\documentclass[sigconf,nonacm]{acmart}

\usepackage{booktabs}
\usepackage{graphicx}
\usepackage{listings}
\usepackage{xcolor}
\usepackage{tikz}
\usetikzlibrary{positioning, arrows.meta, fit, backgrounds}

\renewcommand\footnotetextcopyrightpermission[1]{}
\begin{document}

\title{Workstream: A Local-First Developer Command Center for the AI-Augmented Engineering Workflow}

\author{Happy Bhati}
\affiliation{%
  \institution{Northeastern University}
  \city{Boston}
  \state{Massachusetts}
  \country{USA}
}
\email{bhati.h@northeastern.edu}

\begin{abstract}
Modern software engineers operate across 5--10 disconnected tools daily: GitHub,
GitLab, Jira, Slack, calendar applications, CI dashboards, AI coding assistants,
and container platforms. This fragmentation creates cognitive overhead that
interrupts deep work and delays response to critical engineering signals. We
present \textbf{Workstream}, an open-source, local-first developer command center
that aggregates pull requests, task management, calendar, AI-powered code review,
historical review intelligence, repository AI-readiness scoring, and agent
observability into a single interface. We describe the system architecture, a
novel 5-category AI readiness scoring algorithm, a review intelligence pipeline
that mines historical PR reviews for team-specific patterns, and an agent
observability layer implementing the Model Context Protocol (MCP), Agent-to-Agent
(A2A), and Agent Observability Protocol (AOP). Through a case study of applying
the tool to its own development, we demonstrate measurable improvements in
AI-readiness scores (48 to 98 on our internal scanner; 41.6 to 73.7 on the
independent \texttt{agentready} CLI). Workstream is released as open source under
the Apache 2.0 license at \url{https://github.com/happybhati/workstream}.
\end{abstract}




\maketitle

%% ===================================================================
\section{Introduction}
\label{sec:intro}

The modern software engineer's workday is characterized by relentless context
switching. A 2024 survey by Atlassian found that developers switch between an
average of 9 different tools per day~\cite{atlassian2024}, spending up to 30\%
of their time on ``work about work''---status checks, tool navigation, and
information synthesis---rather than writing code. Research by Mark et
al.~\cite{mark2008cost} demonstrated that interrupted tasks require an average
of 23 minutes and 15 seconds to resume, a finding corroborated by studies
specific to software development~\cite{meyer2014software}.

The emergence of AI coding assistants has intensified this fragmentation. Engineers
now also monitor AI agent status, manage multiple AI providers (Claude, Gemini,
Ollama), track token usage and costs, and assess whether their repositories are
structured to maximize AI agent effectiveness. No existing tool addresses this
full spectrum in a unified interface.

We present \textbf{Workstream}, a local-first developer command center that
consolidates the entire engineering workflow---pull requests, task management,
calendar, AI-powered code review, review intelligence, AI readiness scoring, and
agent observability---into a single dashboard. Our key contributions are:

\begin{enumerate}
\item A \textbf{unified aggregation architecture} that combines data from GitHub,
  GitLab, Jira, and Google Calendar with AI-native features in a local-first,
  zero-configuration deployment model.
\item A \textbf{5-category AI readiness scoring algorithm} that quantitatively
  assesses how well a repository supports AI coding agents, with automated
  gap-filling and draft PR generation.
\item A \textbf{review intelligence pipeline} that mines historical PR reviews to
  extract team-specific patterns and enrich AI code review prompts with
  organizational context.
\item An \textbf{agent observability layer} implementing MCP, A2A, and AOP
  protocols for unified monitoring of AI agents in the developer workflow.
\end{enumerate}

Workstream is implemented in 8,411 lines of Python with a 4,900-line vanilla
JavaScript frontend, supports macOS, Linux, and container deployments, and is
released under the Apache 2.0 license.

%% ===================================================================
\section{Background and Related Work}
\label{sec:related}

\subsection{Developer Productivity Measurement}

The DORA (DevOps Research and Assessment) metrics~\cite{forsgren2018accelerate}
established four key measures of software delivery performance: deployment
frequency, lead time for changes, change failure rate, and time to restore
service. The SPACE framework~\cite{forsgren2021space} broadened this to include
satisfaction, performance, activity, communication, and efficiency. Both
frameworks emphasize reducing friction in the developer workflow, but neither
prescribes specific tooling for unified information aggregation.

\subsection{Existing Developer Dashboards}

Several tools address subsets of the developer information problem:

\begin{itemize}
\item \textbf{GitHub Dashboard}: Shows PRs, issues, and notifications for GitHub
  only. No Jira, calendar, or AI integration.
\item \textbf{Linear}~\cite{linear2024}: Modern project management with GitHub
  integration, but no calendar, AI review, or agent observability.
\item \textbf{Backstage}~\cite{backstage2024}: Spotify's developer portal focuses
  on service catalog and documentation, not daily workflow aggregation.
\item \textbf{Sleuth/Haystack}: DORA-focused dashboards for engineering managers,
  not individual developer workflow tools.
\item \textbf{Raycast/Alfred}: macOS launchers with integrations, but no persistent
  dashboard or AI-native features.
\end{itemize}

None of these tools combine PR management across multiple forges (GitHub + GitLab),
task tracking, calendar, AI code review, review intelligence, AI readiness
assessment, and agent observability in a local-first architecture.

\subsection{AI in Software Engineering}

AI-assisted development has evolved from autocomplete (GitHub
Copilot~\cite{copilot2024}) to agentic coding assistants that perform multi-step
tasks autonomously (Cursor, Claude Code, Codex). This shift creates new needs:
repositories must be structured for AI comprehension~\cite{agentready2024},
AI agents must be monitored and managed, and AI-generated reviews must be
integrated into human workflows with appropriate trust boundaries.

%% ===================================================================
\section{System Design}
\label{sec:design}

\subsection{Design Principles}

Workstream is guided by five principles derived from developer tool research and
practical experience:

\begin{enumerate}
\item \textbf{Local-first}: All data stays on the developer's machine. No cloud
  backend, no account creation, no data sharing. This addresses both privacy
  concerns and organizational security policies.
\item \textbf{Zero-configuration}: A developer can go from \texttt{git clone} to
  a running dashboard with two commands: \texttt{pip install -r requirements.txt}
  and \texttt{python app.py}.
\item \textbf{No build step}: The frontend is a single HTML file with inline CSS
  and JavaScript. This eliminates npm, webpack, and Node.js from the dependency
  chain---a deliberate choice for a tool whose audience is developers.
\item \textbf{Personal instance model}: Each developer runs their own Workstream
  with their own API tokens. There is no multi-tenant mode, no shared database,
  and no token sharing.
\item \textbf{AI-native}: AI features (code review, intelligence, readiness
  scanning, agent monitoring) are first-class citizens, not plugins or
  afterthoughts.
\end{enumerate}

\subsection{Architecture Overview}

\begin{figure*}[ht]
\centering
\begin{tikzpicture}[
  every node/.style={font=\small},
  layer/.style={draw, rounded corners, minimum width=14cm, minimum height=0.9cm,
                fill=gray!8, align=center},
  module/.style={draw, rounded corners, minimum width=2cm, minimum height=0.7cm,
                 fill=white, align=center, font=\small},
  ext/.style={draw, rounded corners, minimum width=2.2cm, minimum height=0.7cm,
              fill=white, align=center, font=\small},
  arr/.style={-{Stealth[length=2.5mm]}, thick},
]

% Browser layer
\node[layer, fill=blue!8] (browser) at (0, 4.5)
  {\textbf{Browser SPA} \quad \texttt{static/index.html} ({\raise.17ex\hbox{$\scriptstyle\sim$}}4{,}900 LOC, vanilla JS, dark/light theme)};

% Arrow
\draw[arr] (0, 4.0) -- node[right, font=\footnotesize] {HTTP/REST + SSE} (0, 3.4);

% FastAPI layer
\node[layer, fill=green!8] (api) at (0, 3.0)
  {\textbf{FastAPI} \quad \texttt{app.py} (38+ endpoints, optional auth middleware)};

% Feature modules
\node[module] (poll) at (-5.5, 1.8) {Pollers};
\node[module] (rev) at (-3.2, 1.8) {AI Reviewer};
\node[module] (rdy) at (-0.7, 1.8) {Readiness};
\node[module] (intel) at (1.8, 1.8) {Intelligence};
\node[module] (agt) at (4.0, 1.8) {Agents};
\node[module] (mcp) at (6.0, 1.8) {MCP Server};

% Background for modules
\begin{scope}[on background layer]
  \node[draw, dashed, rounded corners, fill=yellow!5,
        fit=(poll)(rev)(rdy)(intel)(agt)(mcp),
        inner sep=4pt, label={[font=\footnotesize]above left:Feature Modules}] {};
\end{scope}

% Arrows from API to modules
\foreach \m in {poll, rev, rdy, intel, agt, mcp} {
  \draw[arr, gray] (api.south) -- (\m.north);
}

% SQLite
\node[layer, fill=orange!8, minimum width=14cm] (db) at (0, 0.5)
  {\textbf{SQLite} \quad \texttt{database.py} (aiosqlite, 10+ tables)};

% Arrows from modules to DB
\foreach \m in {poll, rev, rdy, intel, agt, mcp} {
  \draw[arr, gray!60] (\m.south) -- (db.north);
}

% External APIs
\node[ext] (gh) at (-5.0, -0.8) {GitHub API};
\node[ext] (gl) at (-2.5, -0.8) {GitLab API};
\node[ext] (jr) at (0, -0.8) {Jira REST};
\node[ext] (gc) at (2.5, -0.8) {Google Cal};
\node[ext] (ai) at (5.0, -0.8) {AI Providers};

% Arrows from DB to external
\foreach \e in {gh, gl, jr, gc, ai} {
  \draw[arr, gray!60] (db.south) -- (\e.north);
}

\end{tikzpicture}
\Description{Architecture diagram showing five layers: Browser SPA at top,
FastAPI server, six feature modules (Pollers, AI Reviewer, Readiness,
Intelligence, Agents, MCP Server), SQLite database, and five external APIs
(GitHub, GitLab, Jira, Google Calendar, AI Providers) at bottom.}
\caption{Workstream system architecture. The browser communicates with FastAPI
over HTTP and SSE. Six feature modules poll external APIs and persist state in
SQLite. Arrows indicate data flow direction.}
\label{fig:arch}
\end{figure*}
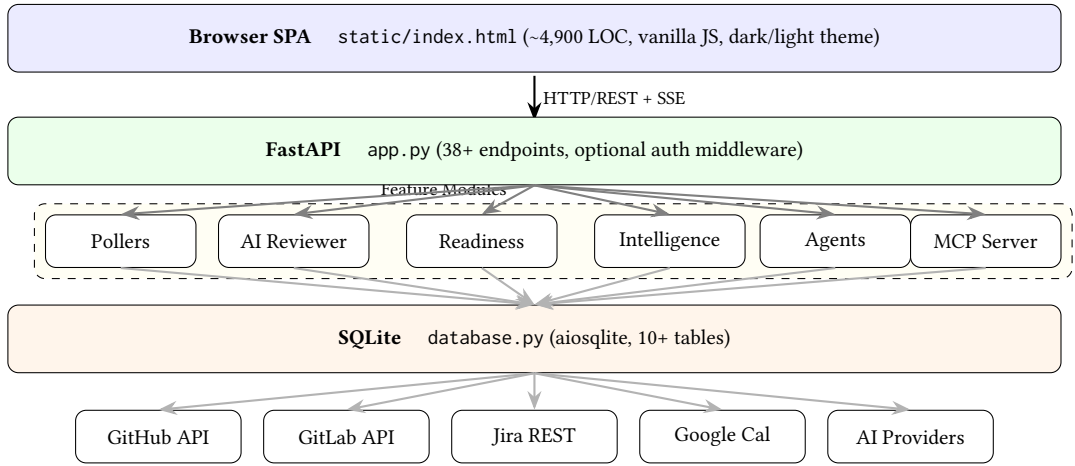

Workstream is a Python FastAPI application serving a single-page frontend and
exposing 38+ REST API endpoints. Figure~\ref{fig:arch} shows the high-level
architecture.

\begin{figure}[ht!]
\centering
\includegraphics[width=\columnwidth]{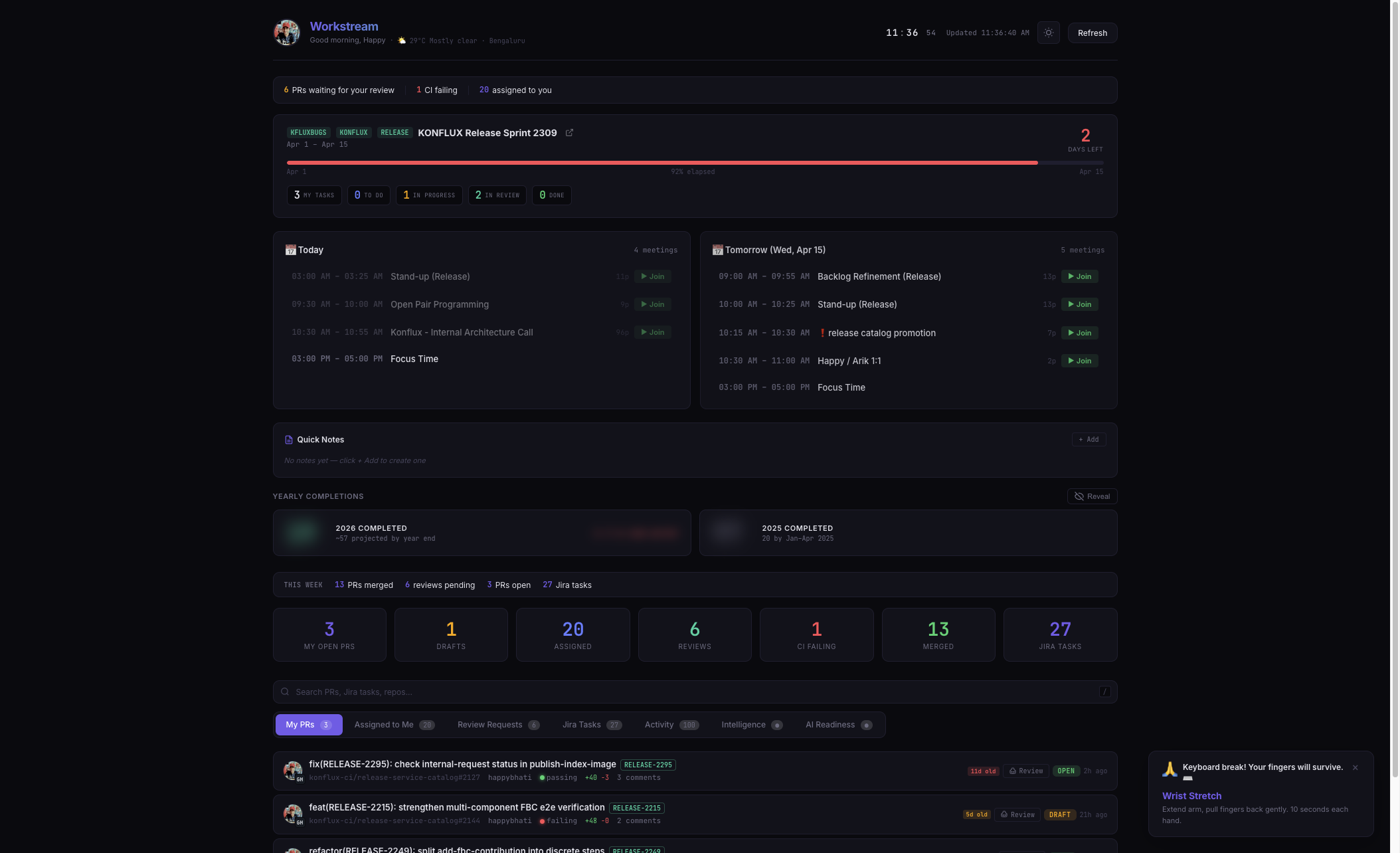}
\Description{Screenshot of the Workstream dashboard showing open pull requests,
calendar events, sprint tracking, and status summary cards in dark mode.}
\caption{Workstream dashboard in dark mode. The single-page interface shows
PRs, calendar, sprint tracking, and status cards simultaneously.}
\label{fig:dashboard}
\end{figure}

The \textbf{polling layer} (\texttt{pollers.py}, 1,132 LOC) runs as an asyncio
background task on a configurable interval (default: 300 seconds). It fetches
data from GitHub (Search API for PRs, Check Runs for CI), GitLab (MR and pipeline
APIs), Jira (REST and Agile APIs for issues and sprints), and Google Calendar
(OAuth2 events API). All fetched data is normalized into SQLite tables with a
unified schema.

The \textbf{frontend} uses CSS custom properties for dark/light theming, tab-based
navigation (My PRs, Assigned, Reviews, Jira, Activity, Intelligence, AI Readiness,
Agents), and direct fetch calls to the REST API. No framework, no virtual DOM, no
build artifacts.

\subsection{Data Model}

The SQLite database contains 10+ tables organized by feature area:

\begin{itemize}
\item \texttt{pull\_requests}: Unified GitHub/GitLab PRs with CI status, reviewers,
  assignees, Jira key, avatars, and \texttt{polled\_at} for stale detection
\item \texttt{activities}: PR events (approvals, change requests, comments)
\item \texttt{jira\_issues}: Cached issues with normalized status categories and
  sprint metadata
\item \texttt{calendar\_events}: Normalized Google Calendar events
\item \texttt{ri\_*}: Review intelligence tables (pull requests, comments, patterns,
  reviewer profiles)
\item \texttt{readiness\_scans}: AI readiness scan history and scores
\item \texttt{agent\_status\_history}, \texttt{agent\_telemetry}: Agent monitoring data
\end{itemize}

\subsection{Security Model}

Workstream handles sensitive data (API tokens, PR content, calendar events) with
a defense-in-depth approach:

\begin{itemize}
\item Tokens are stored only in \texttt{.env} files (git-ignored) and loaded via
  Pydantic settings; they never appear in API responses or frontend config
\item The server binds to \texttt{127.0.0.1} by default, accessible only from
  localhost
\item An optional bearer token gate (\texttt{WORKSTREAM\_AUTH\_TOKEN}) protects
  all endpoints except health checks when network access is needed
\item The AI review engine scrubs secrets (GitHub tokens, AWS keys, PEM blocks,
  Anthropic keys) from PR diffs before sending to any AI provider
\item The container runs as a non-root user with an explicit health check
\end{itemize}

%% ===================================================================
\section{AI-Native Features}
\label{sec:ai}

\subsection{Multi-Provider AI Code Review}

Workstream's review engine (\texttt{reviewer.py}, 591 LOC) supports three AI
providers: Anthropic Claude, Google Gemini, and Ollama (local). The workflow is:

\begin{enumerate}
\item Fetch the unified diff and PR metadata from GitHub or GitLab
\item \textbf{Sanitize}: Regex-based redaction of secrets including patterns for
  \texttt{ghp\_}, \texttt{glpat-}, \texttt{sk-ant-}, \texttt{sk-}, \texttt{AKIA},
  PEM blocks, and Bearer tokens
\item Inject team context from review intelligence (if available for the repository)
\item Send to the selected provider with a structured prompt requesting JSON output
  (summary + file/line/severity comments)
\item Present results in the UI with human approval required before posting
\item On approval, post a consolidated GitHub review or GitLab MR note
\end{enumerate}

The ``copy prompt'' feature allows developers to paste the review prompt into any
LLM not directly integrated, preserving flexibility.

\subsection{Review Intelligence Pipeline}

The intelligence module (\texttt{intelligence/}, 1,509 LOC across 3 files) mines
historical PR reviews to extract organizational knowledge.

\textbf{Collection} (\texttt{collector.py}): Incrementally fetches merged PRs from
the last year via GitHub's API, extracting review bodies and inline review comments.
Comments from bots are filtered. Rate-limit-aware pagination ensures reliable
collection across large repositories.

\textbf{Classification} (\texttt{analyzer.py}): Each review comment is classified
into one of 9 categories using ordered regex patterns:

\begin{table}[ht!]
\caption{Review comment classification taxonomy}
\label{tab:taxonomy}
\begin{tabular}{ll}
\toprule
Category & Example Patterns \\
\midrule
Bug & ``null pointer'', ``race condition'', ``off by one'' \\
Security & ``injection'', ``authentication'', ``CORS'' \\
Error Handling & ``try/catch'', ``error boundary'', ``panic'' \\
Testing & ``unit test'', ``coverage'', ``mock'' \\
Performance & ``O(n\^{}2)'', ``memory leak'', ``cache'' \\
Concurrency & ``deadlock'', ``mutex'', ``thread safety'' \\
API Design & ``REST'', ``backward compatible'', ``breaking'' \\
Documentation & ``docstring'', ``README'', ``comment'' \\
Architecture & ``coupling'', ``dependency'', ``separation'' \\
\bottomrule
\end{tabular}
\end{table}

\textbf{Profile extraction}: For each reviewer, the pipeline aggregates category
distributions, common phrases, and focus areas. For each repository, it extracts
dominant patterns and generates tool-specific insights for AI agents.

\textbf{Integration}: When a developer requests an AI code review, the prompt is
enriched with the repository's intelligence data---dominant review categories,
reviewer preferences, and common patterns---making the AI review contextually
relevant to the team's standards.

\subsection{AI Readiness Scoring Algorithm}

The AI readiness scanner (\texttt{agentic\_readiness/}, 1,654 LOC) evaluates
repositories across five categories using a 120-point rubric normalized to a
0--100 scale.

\begin{table}[ht!]
\caption{AI Readiness scoring rubric}
\label{tab:rubric}
\small
\begin{tabular}{lrp{4.5cm}}
\toprule
Category & Max & Key Signals \\
\midrule
Agent Config & 30 & AGENTS.md, CLAUDE.md, GEMINI.md, .cursor/rules, .codex/config \\
Documentation & 25 & README quality, ARCHITECTURE.md, CONTRIBUTING.md, ADRs \\
CI/CD Quality & 25 & GitHub Actions, linters, pre-commit, Dependabot/Renovate \\
Code Structure & 20 & Modular directories, tests, dependency management \\
Security & 20 & SECURITY.md, .gitignore, no secrets in tree \\
\bottomrule
\end{tabular}
\end{table}

\begin{figure}[ht!]
\centering
\includegraphics[width=\columnwidth]{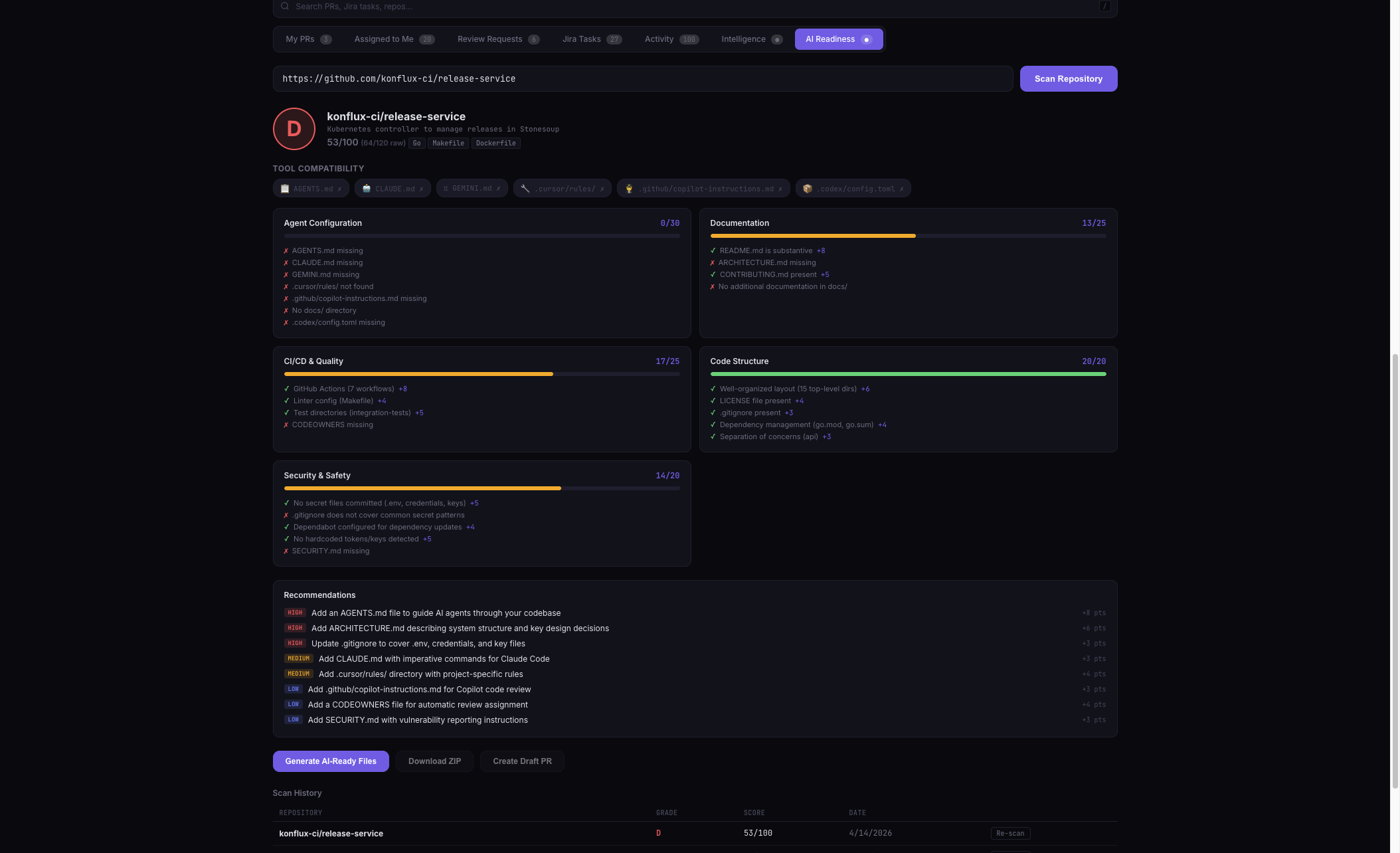}
\Description{Screenshot of the AI Readiness scanner showing a repository score
breakdown across five categories with recommendations.}
\caption{AI Readiness scan results showing the 5-category score breakdown,
letter grade, and prioritized recommendations for improvement.}
\label{fig:readiness}
\end{figure}

The scanner fetches the repository tree, key files, and CI configuration via the
GitHub API and applies heuristic rules for each signal. The scorer produces a
letter grade (A: $\geq$90, B: $\geq$75, C: $\geq$60, D: $\geq$45, F: $<$45)
and generates prioritized recommendations (critical, high, medium, low).

The \textbf{generator} produces gap-filling files: if a repository lacks
\texttt{AGENTS.md}, the generator creates one with the repository map, build
commands, and key conventions extracted from the scan. If review intelligence
exists for the repository, team-specific patterns are injected into the generated
context files.

Generated files can be committed via a \textbf{draft PR} created through the
GitHub Contents API, allowing repository maintainers to review and merge the
AI-readiness improvements through their standard workflow.

\subsection{Agent Observability}

As AI agents become integral to the developer workflow, monitoring them becomes
essential. Workstream implements three emerging protocols:

\textbf{Model Context Protocol (MCP)}: Workstream reads the developer's Cursor
IDE configuration (\texttt{\~{}/.cursor/mcp.json}) to auto-discover configured MCP
servers. For each server, it performs health checks via \texttt{pgrep} (local) or
HTTP probes (remote), tracking status history over time.

\textbf{Agent-to-Agent (A2A)}: Developers can register external A2A agents by
URL. Workstream fetches the agent card from the well-known endpoint
(\texttt{agent-card.json}) and monitors the agent's availability.

\textbf{Agent Observability Protocol (AOP)}: A real-time event bus with
Server-Sent Events (SSE) streaming provides live visibility into agent activity.
Events are categorized as session, cognition, or operation events.

\textbf{Telemetry}: For each agent interaction, Workstream records token usage,
estimated cost (using hardcoded per-model pricing), latency, and success/failure
status. The dashboard displays per-agent and aggregate statistics.

%% ===================================================================
\section{Implementation}
\label{sec:impl}

\subsection{Technology Stack}

Table~\ref{tab:stack} summarizes the technology choices and their rationale.
Table~\ref{tab:metrics} reports the final codebase size.

\begin{table}[ht!]
\caption{Technology stack and rationale}
\label{tab:stack}
\begin{tabular}{lll}
\toprule
Component & Technology & Rationale \\
\midrule
Backend & FastAPI + Uvicorn & Async, auto-OpenAPI \\
Database & SQLite + aiosqlite & Zero-config, local \\
Frontend & Vanilla JS + CSS & No build step \\
Config & Pydantic Settings & Type-safe, env-based \\
AI Providers & httpx & Async HTTP client \\
Container & Podman + Fedora 41 & OCI-compliant, rootless \\
CI/CD & GitHub Actions & Matrix testing, security \\
\bottomrule
\end{tabular}
\end{table}

\subsection{Codebase Metrics}

\begin{table}[ht!]
\caption{Codebase size and composition}
\label{tab:metrics}
\begin{tabular}{lr}
\toprule
Metric & Value \\
\midrule
Python source files & 28 \\
Python lines of code & 8,411 \\
Frontend (HTML/CSS/JS) & {\raise.17ex\hbox{$\scriptstyle\sim$}}4,900 lines \\
REST API endpoints & 38+ \\
Test functions & 33 \\
Test modules & 6 \\
SQLite tables & 10+ \\
External integrations & 7 \\
\bottomrule
\end{tabular}
\end{table}

\subsection{Cross-Platform Deployment}

Workstream supports four deployment models:

\begin{itemize}
\item \textbf{macOS}: LaunchAgent for persistent background service via
  \texttt{install.sh}
\item \textbf{Linux}: Systemd user service via \texttt{bin/workstream-linux}
\item \textbf{Container}: Fedora 41 Podman image with non-root user, published
  to Quay.io
\item \textbf{Kubernetes/OpenShift}: Deployment, Service, and Route manifests
  with Secret templates
\end{itemize}

\subsection{CI/CD Pipeline}

The GitHub Actions pipeline runs five job types:

\begin{enumerate}
\item \textbf{Lint}: Ruff check and format verification
\item \textbf{Test}: Matrix across Python 3.11, 3.12, 3.13 with coverage reporting
\item \textbf{Security}: Dependency audit (\texttt{pip-audit}) and secret scanning
\item \textbf{Container}: Podman build with health check smoke test
\item \textbf{E2E}: Container-based end-to-end test with real API tokens (main
  branch only, using GitHub encrypted secrets)
\end{enumerate}

%% ===================================================================
\section{Evaluation}
\label{sec:eval}

\subsection{Case Study: Dogfooding}

We evaluated Workstream by applying its AI readiness scanner to the Workstream
repository itself and implementing all recommendations. This ``dogfooding''
exercise provides a controlled before-and-after measurement.

\begin{table}[ht!]
\caption{AI Readiness score improvement through dogfooding}
\label{tab:dogfood}
\begin{tabular}{lrrl}
\toprule
Tool & Before & After & Change \\
\midrule
Workstream Scanner & 48/100 & 98/100 & +104\% \\
agentready CLI & 41.6/100 & 73.7/100 & +77\% \\
\bottomrule
\end{tabular}
\end{table}

The improvement was achieved by generating and committing the files recommended
by the scanner: AGENTS.md, CLAUDE.md, GEMINI.md, ARCHITECTURE.md,
CONTRIBUTING.md, SECURITY.md, Cursor rules, Codex config, and five SKILL.md
files following the agentskills.io specification. Additional improvements came
from adding Dependabot configuration, pre-commit hooks, and expanding test
coverage.

The gap between the two tools (98 vs.\ 73.7) reflects differences in their
evaluation criteria. Workstream's scanner emphasizes AI agent context files, while
\texttt{agentready} places greater weight on testing infrastructure and CI
sophistication.

\subsection{Developer Experience Observations}

Through daily use during the development of Workstream itself, we observed several
qualitative improvements:

\textbf{Reduced context switching}: With PR status, Jira tasks, and calendar
visible simultaneously, the need to switch between browser tabs dropped
significantly. The ``focus banner'' provides a real-time summary of what needs
attention, eliminating the mental overhead of synthesizing information from
multiple sources.

\textbf{Faster review response}: Desktop notifications for PR approvals and CI
failures, combined with one-click navigation to the relevant PR, reduced the
time between event and response.

\textbf{AI review adoption}: The ``copy prompt'' feature proved more popular than
direct provider integration, suggesting that developers prefer flexibility in
choosing their AI tool over automated pipelines.

\textbf{Intelligence-enriched reviews}: When review intelligence data was
available for a repository, AI reviews were noticeably more aligned with team
conventions, suggesting that organizational context improves AI output quality.

\subsection{Limitations and Threats to Validity}

\begin{itemize}
\item \textbf{Single-user evaluation}: The observations are from a single developer
  (the author) using the tool during its own development. A controlled study with
  multiple participants would strengthen the evaluation.
\item \textbf{Self-selection bias}: The dogfooding case study measures improvement
  on a repository specifically designed to score well, not on arbitrary repositories.
\item \textbf{Qualitative observations}: The developer experience claims are
  not backed by time-tracking data or controlled experiments.
\item \textbf{API dependency}: Workstream depends on external APIs (GitHub,
  GitLab, Jira) that may change. The polling architecture introduces latency
  compared to webhook-based approaches.
\end{itemize}

%% ===================================================================
\section{Discussion}
\label{sec:discussion}

\subsection{The Developer After AI}

The role of the software engineer is shifting from ``tool operator'' to
``orchestrator.'' Where a developer once wrote code, ran tests, and reviewed diffs
manually, they now coordinate AI agents that draft code, suggest reviews, and
generate documentation. This shift creates a new category of cognitive load:
\textit{agent management overhead}.

Workstream's agent observability layer is an early response to this shift. By
providing a unified view of all AI agents---their status, cost, activity, and
effectiveness---it frees mental bandwidth that would otherwise be spent tracking
agents across disconnected UIs.

\subsection{Dashboard as Cognitive Relief}

The central thesis of Workstream is that \textbf{information consolidation reduces
cognitive load}. When a developer can see their PRs, tasks, calendar, and AI agent
status on a single screen, they spend less mental energy on information retrieval
and more on the engineering work itself.

This is not a new insight---it echoes the ``single pane of glass'' principle from
infrastructure monitoring~\cite{beyer2016site}. What is new is applying it to the
\textit{individual developer's daily workflow} rather than to system operations.

\subsection{Open Questions}

\begin{itemize}
\item \textbf{Multi-user mode}: Can Workstream scale to team dashboards while
  preserving the local-first principle?
\item \textbf{Plugin architecture}: Would a community-driven plugin system for new
  integrations (Slack, Linear, Notion) increase adoption without increasing
  complexity?
\item \textbf{AI readiness benchmarking}: Could the scoring algorithm be validated
  against actual AI agent performance metrics (e.g., task completion rates)?
\end{itemize}

%% ===================================================================
\section{Conclusion}
\label{sec:conclusion}

We presented Workstream, an open-source, local-first developer command center that
addresses the cognitive overhead of modern multi-tool engineering workflows. The
system integrates pull request management, task tracking, calendar, AI-powered code
review with secret scrubbing, historical review intelligence, a quantitative AI
readiness scoring algorithm, and agent observability across MCP, A2A, and AOP
protocols.

Through a case study of dogfooding on the Workstream repository itself, we
demonstrated that the AI readiness scoring and gap-filling pipeline can
substantially improve a repository's readiness for AI-augmented development (48 to
98 on our scanner, 41.6 to 73.7 on the independent agentready CLI).

Workstream represents a step toward rethinking the developer experience in the
AI era: rather than adding more tools, we consolidate existing ones and make AI a
first-class participant in the workflow. The project is available at
\url{https://github.com/happybhati/workstream} under the Apache 2.0 license.

%% ===================================================================
\section*{Acknowledgments}

The author thanks the open-source communities behind FastAPI, SQLite, and the
emerging AI agent protocols (MCP, A2A, AOP) that made this work possible.

%% ===================================================================
\bibliographystyle{ACM-Reference-Format}

\end{document}